\documentclass[prl,
 reprint,
 amsmath,amssymb,
 aps,
floatfix,
]{revtex4-2}

\usepackage{graphicx}
\usepackage{dcolumn}
\usepackage{bm}
\usepackage{physics2}
\usephysicsmodule{ab}   
\usephysicsmodule{ab.braket} 
\usepackage{fixdif, derivative} 

\usepackage[whole]{bxcjkjatype} 

\usepackage{color}

\begin{document}

\preprint{APS/123-QED}

\title{Quantum Optical Spanner: \\Twisting Superconductors with Vortex Beam via Higgs Mode}

\author{Daemo Kang}
\author{Sota Kitamura}

\author{Takahiro Morimoto}%

\affiliation{Department of Applied Physics, The University of Tokyo, Hongo, Tokyo,
  113-8656, Japan}
%

\date{\today}

\begin{abstract}
Light carrying orbital angular momentum (OAM)—known as vortex beams—has broadened the scope of understanding and applications of light's angular momentum. Optical tweezers using OAM, often referred to as optical spanners, have significantly expanded the tunability of optical manipulation. A key frontier now lies in understanding how vortex beams interact with quantum states of matter. In this work, we numerically investigate the dynamics of a superconductor under vortex beam illumination and demonstrate the transfer of angular momentum from light to the superconducting collective mode, resulting in mechanical rotation. Our findings open a
pathway for optical manipulation in the quantum regime, which we term the  \textit{quantum optical spanner}.
\end{abstract}

\maketitle


\textit{Introduction. ---}
The development of laser technology has progressed remarkably over the past few decades. Especially, the introduction of light carrying orbital angular momentum (OAM) by Allen et al.~\cite{PhysRevA.45.8185} has brought major advances in optical technology. Vortex beams, light carrying OAM, lead to wide-ranging applications in optical technology, quantum computing\cite{mair2001entanglement, nagali2009quantum}, optical communications\cite{wang2012terabit}, radar imaging and detection \cite{lavery2013detection}, and optical tweezers\cite{he1995direct,grier2003revolution,padgett2011tweezers, kuga1997novel,o2002intrinsic,macdonald2002creation}.

The incorporation of OAM into optical manipulation has led to the development of so-called ``optical spanners,'' which induce rotation in trapped particles through angular momentum transfer~\cite{he1995direct}.  With recent advances in beam shaping and tunable light sources, the control of trapped particles has become increasingly precise and versatile~\cite{padgett2011tweezers,grier2003revolution}. In addition, metal particles can also be manipulated by plasmonic tweezers, demonstrating its flexibility to the nanoscale~\cite{Shen:12,zhang2015plasmonic}.

The generation of vortex beams spans a broad range of frequencies, including the optical, microwave, terahertz (THz), and even X-ray regimes~\cite{wang2018recent,zhang2020review,zhu2019review,chen2018orbital}.  In the microwave regime, vortex beams have been shown to induce orbital currents and Kapitza stabilization in superconducting circuits \cite{PhysRevB.110.144519}. Advances in the THz domain are especially relevant to condensed matter physics~\cite{Xie:2013aa,he2013generation, RevModPhys.94.035003}. In this regime, vortex beams can transfer their spatial phase structure to matter, imprinting the OAM momentum patterns in magnetic materials~\cite{PhysRevB.96.060407}. More broadly, such structured light has been proposed as a tool for controlling topological excitations, including the creation and manipulation of magnetic and polar skyrmions~\cite{PhysRevB.95.054421, PhysRevLett.132.026902}, as well as light-induced vortices in superconductors~\cite{doi:10.7566/JPSJ.89.103703, yeh2025structuredlightinducedvorticityI,yeh2025structuredlightinducedvorticity}.

Vortex beams in the THz range have spatial profiles much larger than atomic lattice constants, making it difficult for their OAM to directly influence electronic quantum states within the diffraction limit. However, superconductivity is a macroscopic quantum phenomenon, and thus can interact with the helical phase structure of such light. This interaction has led to theoretical proposals such as the excitation of spiral-shaped Higgs modes and the enhancement of third-harmonic generation (THG)~\cite{PhysRevResearch.5.L042004}. Despite these advances, a fundamental question, how optical OAM transforms into mechanical angular momentum in quantum matter, has not been addressed yet. 
Elucidating superconductor dynamics under vortex beam would represent a direct demonstration of OAM transfer from light to quantum matter, analogous to Beth's early experiment on spin angular momentum transfer and the work of He et al. on OAM transfer~\cite{he1995direct, PhysRev.50.115}.

We investigate the time evolution of angular momentum in a superconductor under irradiation by a vortex beam by numerically solving the time-dependent Ginzburg--Landau (TDGL) equation. We reveal that a vortex beam can efficiently transfer its OAM to the superconductor via the excitation of Higgs modes. Remarkably, our order estimates indicate that the resulting angular momentum transfer is sufficiently large to generate a mechanical rotation that is experimentally feasible to probe. 
This reveals a fundamentally distinct mechanism for twisting bulk quantum matter, contrasting with earlier approaches based on classical electrodynamics. Unlike conventional optical forces that act on individual particles, our mechanism relies on the quantum-coherent dynamics of the superconducting order parameter, enabling macroscopic angular momentum transfer via vortex beams. These findings open a pathway for non-contact manipulation of quantum matter using the OAM of light in the quantum regime---realizing a \textit{quantum optical spanner}.

\begin{figure}[t]
\includegraphics[width=\columnwidth]{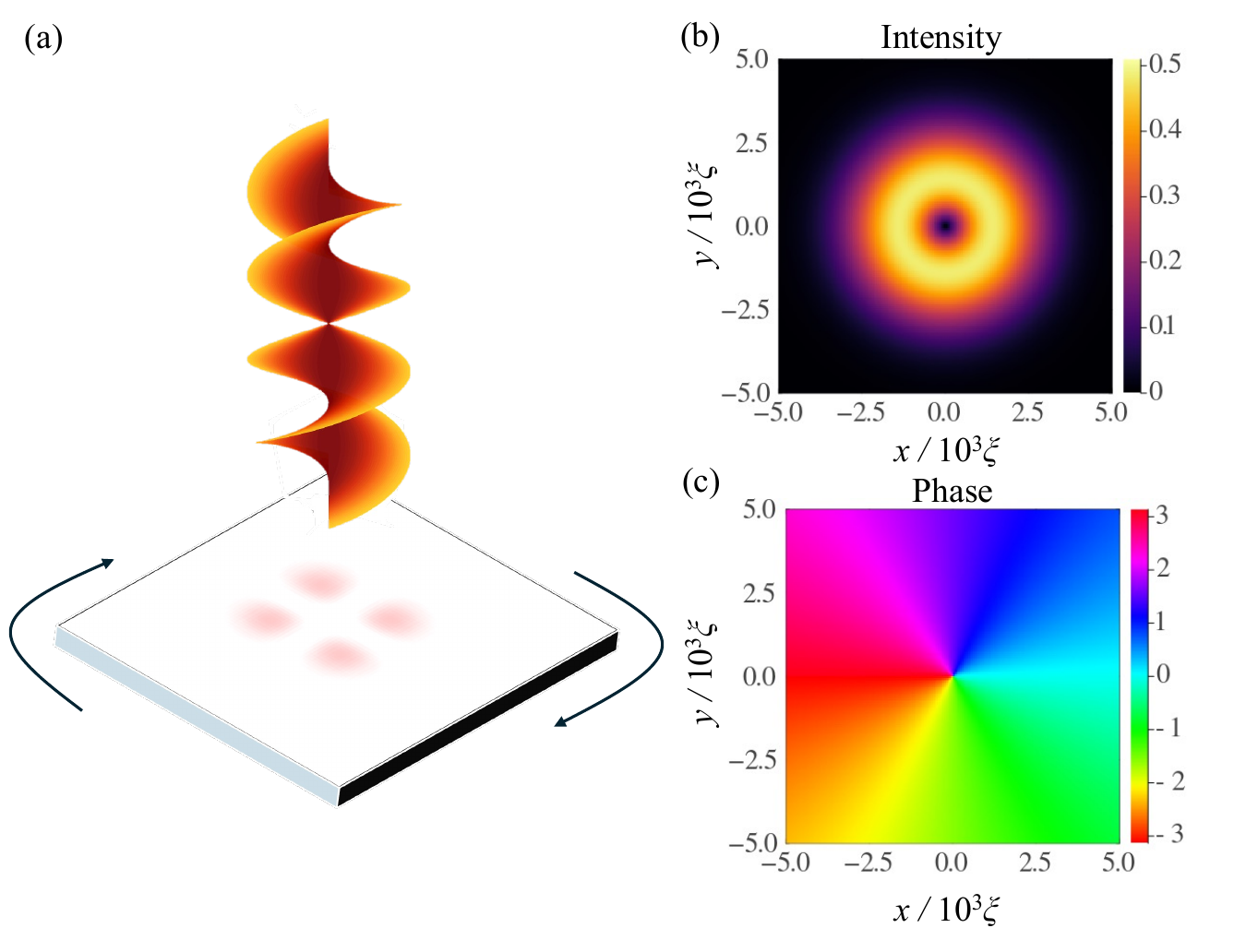}
\caption{\label{fig1} 
(a) Schematic illustration of angular momentum transfer to a superconductor via vortex beam-induced Higgs mode excitation, observable as mechanical rotation. 
(b, c) Real space intensity and phase profiles of the LG beam with OAM $l=1$.}
\end{figure}

\textit{Formalism. ---}
Laguerre-Gaussian (LG) mode is a class of solutions of Helmholtz equation in a cylindrical coordinate system with finite OAM \cite{PhysRevA.45.8185}. As shown in Fig.~\ref{fig1}, the LG beams exhibit a ring-shaped intensity distribution and a helical phase front, with a phase singularity at the center on its cylindrical axis. The form of LG mode at $z=0$, $u_{l,p}(r,\phi, z=0)$ is expressed as
\begin{equation}       
     u_{l,p}(r,\phi) =  \sqrt{\frac{2p!}{\pi(p+|l|)!}}\left(\frac{\sqrt{2}r} {w_0}\right)^{|l|} e^{-\frac{r^2}{w_0^2}}e^{il\phi}L^{|l|}_p\left(\frac{2r^2}{w_0^2}\right)
\end{equation}
Here, $r,\phi,z$ denote the radial, azimuthal, and axial coordinates. The integers $l$ and $p$ specify the mode of the generalized Laguerre polynomial $L^p_{|l|}$. Specifically, $l$ is the azimuthal index determining the OAM, and $p$ controls the number of radial nodes. 
The beam width is set as the beam waist $w_0$ at the focal position $z=0$.
The vortex beam with finite OAM is realized by LG modes with $l \neq 0$, while $(l, p) =(0,0)$ corresponds to the conventional Gaussian beam without OAM.

We consider an $s$-wave superconductor driven by the vortex beam.
We employ the Ginzburg-Landau (GL) theory to study the dynamics of a superconducting order.The GL Lagrangian with the complex order parameter $\Psi(\bm{r})$ is given by \cite{tinkham2004introduction,TSUJI2024174}
\begin{align}\label{Lagrangian}
  \mathcal{L} =& -\left[a|\Psi|^2+\frac{b}{2}|\Psi|^4+\frac{1}{2m^{*}}\left|\left(-i\hbar\nabla-e^{*} \bm{A}\right)\Psi\right|^2\right], \nonumber\\
    &+c\left|i\hbar\frac{\partial \Psi}{\partial t}\right|^2 +d\Psi^{\ast}i\hbar\frac{\partial \Psi}{\partial t},
\end{align}
with the material-dependent parameters $a,b,c$, and $d$.
Here, $a$ is expressed as $a = a_0(T-T_c)$ with $a_0>0$, near the critical temperature $T_c$. 
The equilibrium order parameter $\Psi_0$ in the equilibrium is given by $\Psi_0=\sqrt{-a/b}$ with $b>0$, leading to superfluid density $n_s =|\Psi_0|^2=-a/b$. The coefficients $c$ and $d$ govern the kinetic dynamics of the system. Hereafter we choose $d=0$ because of the particle-hole symmetry\cite{TSUJI2024174, PhysRevB.92.064508}.
$e^{*}$ and $m^{*}$ are the charge and mass of a Cooper pair.
We introduce the LG beam irradiation as a vector potential $\bm{A} = \textrm{Re}\ab[A_0\bm{n}_su_{l,p}(r,\phi)\exp(-i\Omega t)f(t)]$ with $\bm{n}_s$ representing the polarization. We consider linearly polarized light and set $\bm{n}_s=\bm{e}_x$. The envelope function $f(t)$ determines the temporal profile of the input.

The equation of motion for the order parameter $\Psi$ obtained from Eq.(\ref{Lagrangian})
gives the TDGL equation, 
\begin{equation}
  c\hbar^2\frac{\partial^2 \Psi}{\partial t^2} +  \Gamma\pdv{\Psi}{t} = -\ab[a\Psi +  b\Psi|{\Psi}|^2 +\frac{1}{2m^*}(-i\hbar\nabla-e^{*}\bm{A})^2\Psi ] 
\end{equation}
Here, we introduced the dissipation term $\Gamma \partial\Psi/\partial{t}$ phenomenologically. This term does not conserve energy, so it does not change the form of the conserved quantities.
For the numerical computation of the TDGL equation, we employ the dimensionless form, 
\begin{equation}\label{eq:nondimTDGL}
    \tau\frac{\partial^2 \psi}{\partial t'^2} + \gamma\pdv{\psi}{t'} =  \psi - \psi|\psi|^2 + (\nabla'- i\bm{A}')^2 \psi,
\end{equation}
by using the following rescaling of the parameters: $\psi = \Psi/\Psi_0$, $\tau=1/2$,  $t' =  \Omega_{\text{Higgs}}t$, $\gamma=\Gamma\Omega_{\text{Higgs}}/(-a)=0.1$ with the Higgs mode frequency $\hbar\Omega_{\text{Higgs}}=\sqrt{-a/2c}=\Delta_0$ (the superconducting gap), $\bm{r}'=\bm{r} / \xi$ with the coherence length $\xi= \sqrt{\hbar^2/4m_e(-a)}$, and $\bm{A}' = \bm{A}/A_0$ with $A_0 = \hbar/(2e\xi)$.
Here, the Higgs mode is a collective excitation of amplitude of the order parameter ($\delta \Psi$) and is coupled to the square of the electromagnetic field $A^2$ \cite{PhysRevB.92.064508}. 
We note that the present treatment with the TDGL equation 
gives a phenomenological approach to explore the fast dynamics of the superconducting order parameter compared to the relaxation processes, particularly on time scales comparable to the oscillation frequency of the Higgs mode\cite{PhysRevB.93.014503}.

The angular momentum of a superconductor is obtained within the GL description as a Noether charge $L_z = \int dr L_z(\bm{r})$ with the angular momentum density $\epsilon L_z(\bm{r}) = (\delta\mathcal{L}/\delta\dot{\Psi})\delta\Psi$ under a rotational transformation, given by $\delta\Psi = \epsilon(\bm{r}\times\nabla \Psi)$. 
Specifically, the angular momentum density is written as 
\begin{equation}\label{eq:angular_momnetum}
  L_z (\bm{r}) = c\hbar^2\dot{\Psi}^*(\bm{r}) \bm{r} \times \nabla \Psi(\bm{r}) + h.c. 
\end{equation}
In the following, we handle the dimensionless form of the angular momentum $L_z'$ defined by 
\begin{equation}
    L_z'=\frac{L_z}{c(\Omega_{\text{Higgs}}/{2\pi})\hbar^2n_s\xi^3}.
\end{equation}

\begin{figure}
\includegraphics[width=\columnwidth]{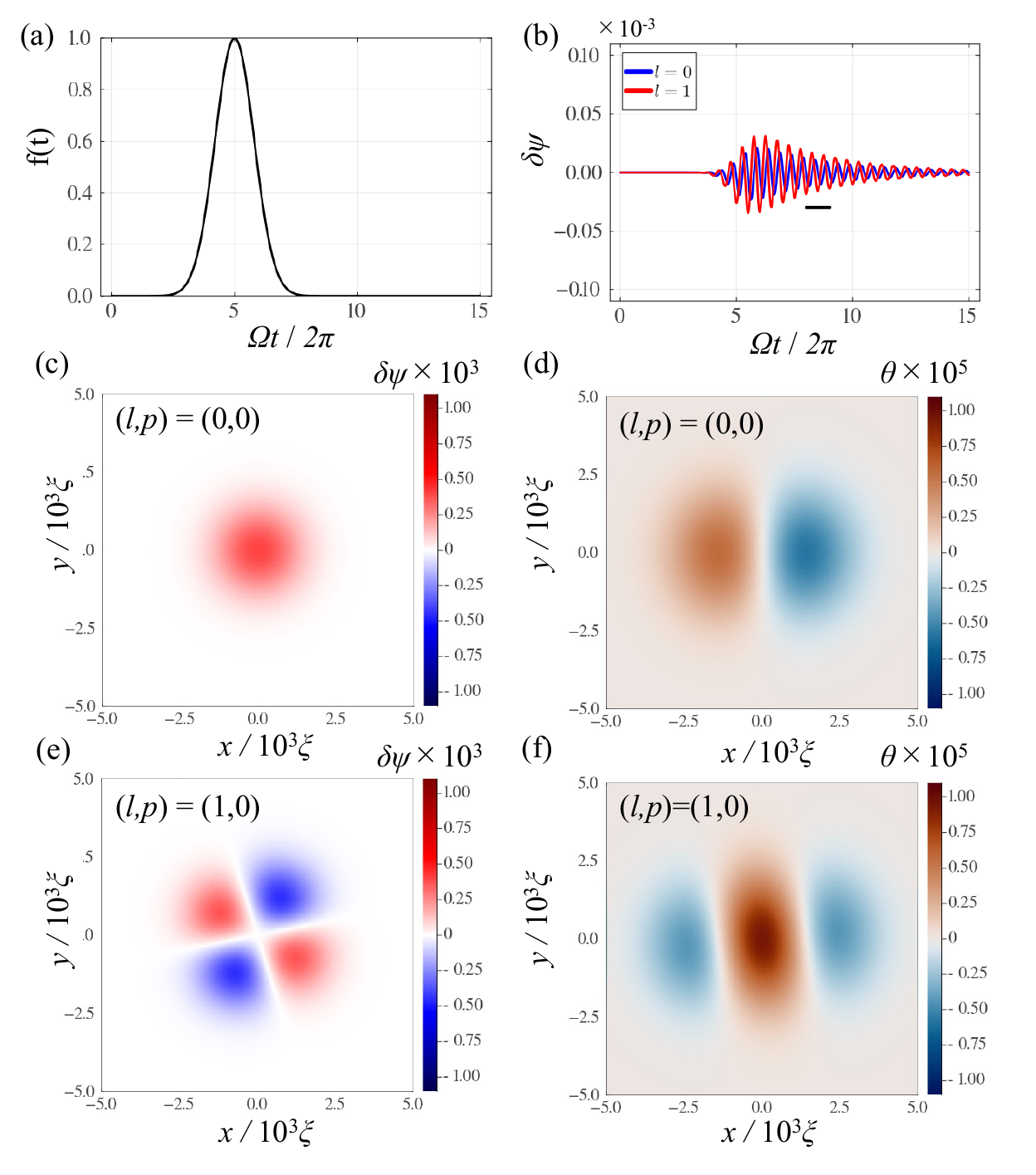}
\caption{\label{fig2} 
Dynamics of the superconducting order parameter under vortex beam irradiation.
(a) Envelope function of the beam, which peaks at $\Omega t/ 2\pi = 5$. 
(b) Time evolution of the amplitude modulation $\delta \psi$ at $(x,y)=(1.25,1.25)/10^3\xi$ for both Gaussian beam and vortex beam. The time scale bar (black line) corresponds to one period of the beam.
(c-f) Amplitude modulation $\delta \psi(\bm{r},t)$ and phase $\theta(\bm{r},t)$ of the order parameter at $\Omega t/ 2\pi = 6$. (c) and (d) show the results under the Gaussian beam $(l,p) = (0,0)$ irradiation, while (d) and (e) show the results under vortex beam $(l,p)=(1,0)$ irradiation.}
\end{figure}

\textit{Result. ---}
We investigate the time evolution of the superconducting order parameter by numerically solving the TDGL equation. 
We consider a Gaussian pulse of the vortex given by the envelope function 
$f(t) = \exp(-(t-t_0)^2 / (2\sigma^2))$ with $\sigma = 10\Omega$, $t_0 = 10\pi/\Omega$ as shown in Fig.~\ref{fig2}(a). 
In the simulations, the beam waist is set to $w_0 = 2000\xi$, and the system size is $10000\xi$, corresponding to the spatial domain $x,y \in [-5000\xi, 5000\xi]$. 
Since the typical coherence length of a superconductor is of the order of \(\xi \sim \mathcal{O}(100\,\mathrm{nm})\), the present beam waist corresponds to $w_0 \sim 200\,\mathrm{\textmu m}$, which lies within the diffraction limit for a 1THz beam and is experimentally accessible. Here, the input frequency is set to 
$ \hbar\Omega =\Delta_0$, which is at the Higgs mode resonance arising from the two-photon absorption. The amplitude of the light is set to \(|A|/A_0 =0.01\).

The superconducting order parameter is expressed as $\psi = (|\psi|+\delta\psi)e^{i\theta}$, and we analyze the amplitude modulation $\delta\psi$ and the phase $\theta$ for $l=0$ and $l=1$ in Fig.~\ref{fig2}. 
For both Gaussian and vortex beams, the order parameter at a given site exhibits oscillations at twice the frequency of the external electric field [Fig.~\ref{fig2} (b)]. After the external field is turned off, these oscillations gradually decay due to the dissipation introduced by the $\gamma$ term, 
which describes the damping of the Higgs mode excitation.
Figures~\ref{fig2}(c--e) show the spatial distributions of the amplitude and phase of the order parameter under Gaussian and vortex beam irradiation. As seen in Figs.~\ref{fig2}(c) and (e), irradiation with Gaussian and vortex beams induces amplitude fluctuations whose spatial profiles follow the spatial structure of the vector potential $\bm{A}^2$, resulting in a two-fold rotational symmetry due to their coupling (the spatial pattern of the vector potential $\bm{A}$ is shown in Supplemental Material). In contrast, the phase distributions in Figs.~\ref{fig2}(d) and (f) reflect the spatial pattern of $i \partial_x A_x$ appearing in the kinetic term of the TDGL equation~(\ref{eq:nondimTDGL}). Unlike the amplitude fluctuations, the phase fluctuations do not exhibit any rotational dynamics after the pulse ends.

\begin{figure}[tb!] 
  \centering
  \includegraphics[width=\columnwidth]{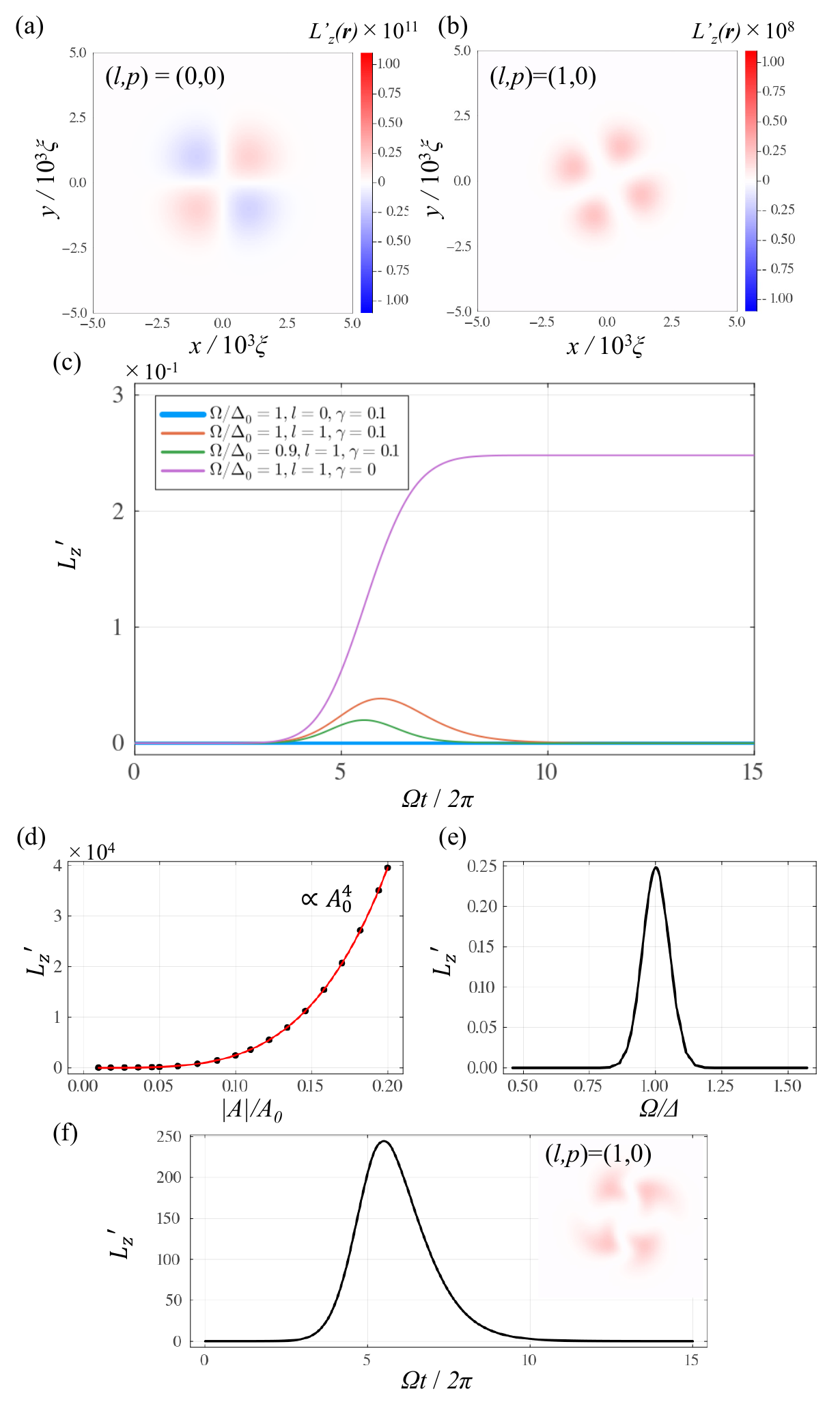}
  \caption{\label{fig3} The time evolution of angular momentum transfer. 
  (a,b) Angular momentum density $L_z(\bm{r})$ at $\Omega t/2\pi= 7.6 $ under (a) the Gaussian beam $(l,p)=(0,0)$ and (b) the vortex beam $(l,p)=(1,0)$. (c) Time evolution of the total angular momentum $L_z$ of the superconductor under Gaussian ($l=0$) and vortex ($l=1$) beams. The incident beam is resonant with the Higgs mode for $\Omega/\Delta_0=1$ and off-resonant for $\Omega/\Delta_0=0.9$, where results with dissipation ($\gamma=0.1$) as well as without dissipation are shown.  (d,e) Dependence of angular momentum transfer on (d) the amplitude $A_0$ and (e) the frequency $\Omega$ of the incident beam, calculated for the case without dissipation ($\gamma=0$). (f) Time evolution of the angular momentum transfer obtained with a smaller sample size $100\xi$, a beam spot $w_0=20\xi$ and a stronger amplitude $|A|/A_0=1$. The inset is for the corresponding spatial pattern of $L_z(\mathbf{r})$.
  }
  \label{fig:wide}
\end{figure}

Figure \ref{fig3} shows the dynamics of the angular momentum transfer. 
Under Gaussian beam irradiation, the angular momentum density exhibits a symmetric pattern with positive (blue) and negative (red) components oscillating around zero, as shown in Fig.~\ref{fig3}(a). In contrast, under vortex beam irradiation (Fig.~\ref{fig3}(b)), the angular momentum density distribution evolves asymmetrically in sign: it predominantly exhibits components of a single sign (e.g., only red regions), indicating a net angular momentum transfer to the superconductor from the beam's angular momentum.

The total angular momentum remains zero when a Gaussian beam is applied, while
nonzero angular momentum transfer occurs when a vortex beam is applied (Fig.~\ref{fig3}(c)). Furthermore, when the incident light frequency is close to the Higgs mode oscillation frequency, the transferred angular momentum remains even after irradiation,
since the Higgs mode is a real excitation of the amplitude $\delta \psi$ that persists even after the pulse ends. 
When dissipation is included ($\gamma>0$), the oscillations of the order parameter are gradually damped.
In this process, the angular momentum imprinted onto the superconductor is transferred to other degrees of freedom such as quasiparticles, phonons, and heat. Namely, this damping process provides a channel through which the light-induced angular momentum induces a mechanical rotation of the superconductor in the end.
In addition, for off-resonant frequencies, the angular momentum transfer is reduced, since the rotational excitation of the amplitude $\delta \psi $ is suppressed due to the detuning, resulting in weaker coupling to the Higgs mode.
Figure \ref{fig3}(d) shows that this angular momentum transfer increases with the fourth power of the electric field. 
This behavior arises from the two-photon absorption of Higgs mode resonance, where the energy absorption is $\propto E^4$, and hence the angular momentum transfer also scales as $\propto E^4$. 
As demonstrated in Fig. \ref{fig3}(e), the angular momentum transfer exhibits a Lorentzian form with a peak at the Higgs mode resonance $\Omega/\Delta_0=1$, 
showing that the transfer of optical OAM to the superconductor is efficiently induced by Higgs mode excitation. Figure~\ref{fig3}(f) shows results obtained for a smaller sample, beam waist and a stronger field amplitude, used for order-of-magnitude estimation. The panel presents both the time evolution of the angular momentum transfer and the corresponding spatial profile $L_z(\mathbf{r})$, demonstrating that the essential features of the dynamics are preserved even under these conditions with the strong light field.

\begin{figure}
\includegraphics[width=0.8\columnwidth]{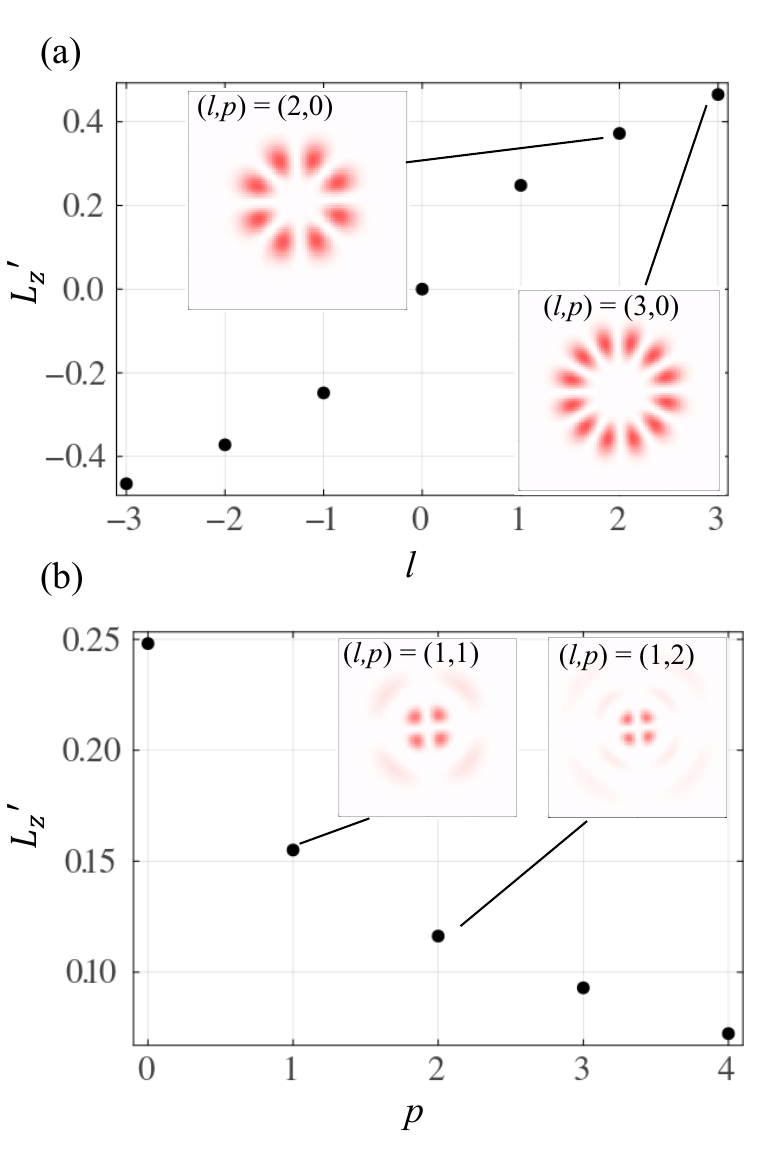}
\caption{\label{fig4} 
Angular momentum of the superconductor as a function of the vortex beam parameters, calculated in the absence of dissipation ($\gamma=0$)
(a) Angular momentum transfer as a function of the beam's angular momentum $l$. Insets show spatial patterns of vortex beam for $(l,p)=(2,0)$ and $(l,p)=(3,0)$.
(c) Angular momentum transfer as a function of the radial index $p$.
Insets show spatial patterns of vortex beam for $(l,p)=(1,1)$ and $(l,p)=(1,2)$.}
\end{figure}

Figure~\ref{fig4} illustrates how different types of the LG beam affect the transfer of angular momentum, obtained without dissipation $(\gamma=0)$.
Figure~\ref{fig4}(a) shows the dependence of $L_z$ on the OAM \( l \) of the LG beam. 
One can see that the angular momentum transfer increases almost proportionally with the OAM \( l \) of the LG beam as expected. 
Also, we find that the number of spatial regions with finite angular momentum density increases with \( l \).
Figure~\ref{fig4}(b) shows the dependence of $L_z$ on the radial index \( p \), which does not affect the total angular momentum of the light but introduces concentric phase singularities. 
As \( p \) increases, the angular momentum transfer decreases due to partial cancellation.
Since the LG beam has a phase reversal across the phase singularities with higher $p$, opposite angular momentum density appears across those phase singularities, resulting in a partial cancellation of the total angular momentum.

\textit{Discussion. ---}
To assess the feasibility of angular momentum transfer, we estimate the relevant quantities using parameters for an $s$-wave superconductor. The total angular momentum of the superconductor associated with the order parameter is given by $L_z = 
\int d\bm{r}^3 L_z(\bm{r}) = \int d\bm{r}^3c\hbar^2\dot{\Psi}(\bm{r}\times\nabla\Psi(\bm{r}))$ which corresponds to the mechanical angular momentum of the superconductor expressed as $ L_z = I\Omega_s$ where \( I = \rho h d^4 / 6 \) is the moment of inertia of a square thin film of side length \( d \), thickness \( h \), and mass density \( \rho \), and \( \Omega_s \) is the resulting rotational frequency.
From these two equations, the rotational frequency of the superconductor induced by vortex beam irradiation can be written as
\begin{equation}\label{eq: rotation}
    \Omega_s = \frac{\hbar^2}{4\pi}\frac{n_s}{\Delta_{0} I}L_z'.
\end{equation}

A larger angular momentum transfer is expected in superconductors characterized by a small energy gap, a high superconducting carrier density, and a low mass density.  
Here, we consider niobium nitride (NbN) as a typical example of an \( s \)-wave superconductor. NbN has a superconducting gap of \( 2\Delta_0 = 2.5~\mathrm{meV} \), a coherence length of as large as \( \xi = 4~\mathrm{nm} \), and a superconducting carrier density of \( n_c = 6 \times 10^{28}~\mathrm{m}^{-3} \)\cite{doi:10.1126/science.1254697}. This value corresponds to the total conduction electron density at low temperatures, assuming that nearly all conduction electrons form Cooper pairs in the superconducting state.
In this case, the Higgs resonance frequency of light is estimated to be \( \Omega = 1.8 \times 10^{12}~\mathrm{Hz} \).
We consider a square thin film sample with dimensions $10000\xi \times 10000\xi \times 10\xi  = 40~\mathrm{\mu m} \times 40~\mathrm{\mu m}\times 40~\mathrm{nm} $.  
The typical value of the angular momentum \(L_z' = 2.4\times10^{-1}\) in Fig.~\ref{fig3} leads to the resulting rotational frequency of \( \Omega_s \sim 2.3 \times 10^{-20} \)~rad/s.  

While the present choice of the field strength $|A|/A_{0}=0.01$ (corresponding to an electric field of  $1.5 \times 10^{-2}$ kV/cm) renders the induced rotational frequency $\Omega_{s}\sim 2.3\times10^{-20}$ rad/s far too small to observe, a stronger field and smaller sample size lead to experimentally accessible values. For instance, if we adopt $|A|/A_{0}=1.0$ and consider a square thin film of niobium nitride with dimensions $100\xi\times100\xi\times10\xi=0.4~\mu\mathrm{m}\times0.4~\mu\mathrm{m}\times40~\mathrm{nm}$ and a beam waist $w_0=20\xi=80~\mathrm{nm}$ \cite{guo2024terahertz}, the required field strength is about $1.5$ kV/cm. This results in an angular velocity of $5.9\times10^{-8}$ rad/s by one pulse, using the peak value in Fig.~\ref{fig3}(f). Using a pulsed laser with a continuous 100 MHz pulse train, this leads to several rotations within a few seconds, confirming that the effect is experimentally feasible.

\begin{acknowledgments}
We thank K. Tanaka, R. Shimano and H. Watanabe for fruitful discussions.
This work was supported by JST CREST, Grant Number JPMJCR19T3, JSPS KAKENHI Grant 23K25816, 23K17665, 24H02231 (T.M.), and JSPS KAKENHI Grant 25K07219 (S.K.). 
D.K. was supported by the Forefront Physics and
Mathematics program to drive transformation (FoPM).

\end{acknowledgments}

\bibliographystyle{apsrev4-1}
\bibliography{paper}

\end{document}


\preprint{APS/123-QED}

\title{Supplemental Material for ``Quantum Optical Spanner: \\Twisting Superconductors with Vortex Beam via Higgs Mode"}

\author{Daemo Kang}
\author{Sota Kitamura}

\author{Takahiro Morimoto}%

\affiliation{Department of Applied Physics, The University of Tokyo, Hongo, Tokyo,
  113-8656, Japan}
%

\date{\today}

\maketitle


\section{Spatial Profiles of Gaussian and Laguerre-Gaussian Beams}

\begin{figure}[tb] 
  \centering
  \includegraphics[width=\columnwidth]{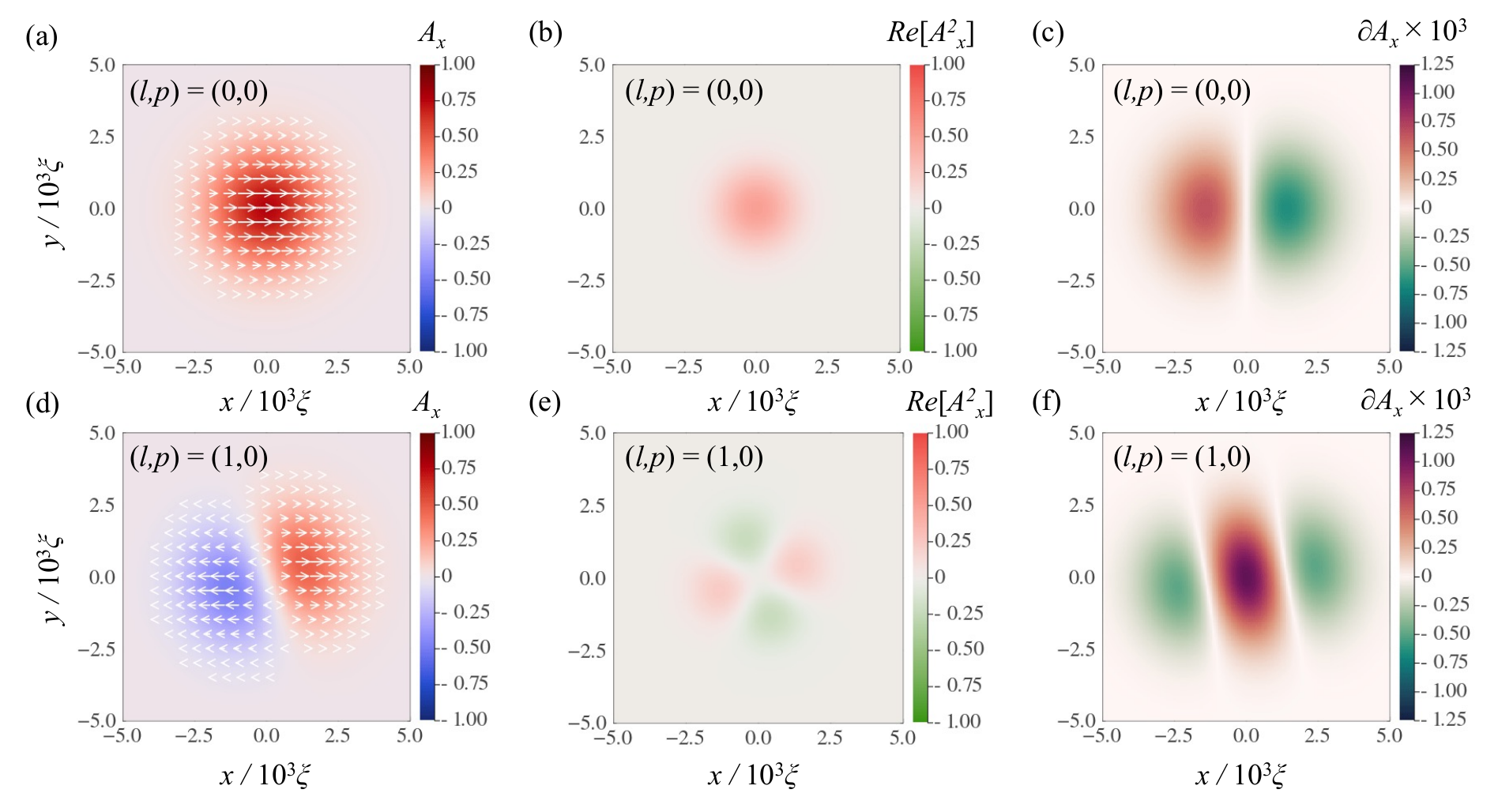}
  \caption{\label{SupplementalMaterial} Field profiles and derived quantities under Gaussian and Laguerre–Gaussian (LG) beams at $\Omega t /2\pi=6$.
  Panels (a–c) correspond to the Gaussian beam with $(s,l,p)=(0,0,0)$:
  (a) vector potential $A_x$ shown as a background color plot together with the vector field $\vec{A}$,
  (b) real part of the squared vector potential, $\mathrm{Re}[A_x^2]$,
  and (c) spatial derivative $\partial_x A_x$.
  Panels (d–f) correspond to the LG beam with $(s,l,p)=(0,1,0)$:
  (d) $A_x$ with the overlaid vector field $\vec{A}$,
  (e) $\mathrm{Re}[A_x^2]$,
  and (f) $\partial_x A_x$.
  }
  \label{fig:wide}
\end{figure}

Here we show the spatial profiles of Gaussian and Laguerre-Gaussian beams employed in the present study. The vector potential is given by
\begin{equation}
  \vec{A}_{s,l,p}(\mathbf{r},t) = 
  \mathrm{Re}\!\left[A_0\,\hat{n}_s\,u_{l,p}(r,\phi)\,e^{-i\Omega t}f(t)\right],
\end{equation}
where $A_0$ is the amplitude, $\hat{n}_s$ the polarization vector, 
and $u_{l,p}(r,\phi)$ the transverse mode profile [Gaussian for $(l,p)=(0,0)$, LG otherwise]. 
Since in Fig.~\ref{SupplementalMaterial} we focus solely on the spatial profile, 
we set $A_0=1$ and $f(t)=1$, such that
\begin{equation}
  \vec{A}_{s,l,p}(\mathbf{r}) \propto \hat{n}_s\, u_{l,p}(r,\phi), 
  \qquad 
  A_x(\mathbf{r}) \propto u_{l,p}(r,\phi), \quad A_y(\mathbf{r})=0
\end{equation}
for the case of $x$-polarized beams ($s=0$).

To clarify how the spatial structure of the external field governs the induced modulation of the superconducting order parameter, Fig.~\ref{SupplementalMaterial} shows three representative quantities: (a–c) Gaussian beam with $(s,l,p)=(0,0,0)$, and (d–f) LG beam with $(s,l,p)=(0,1,0)$. 
The figure shows $A_x(\mathbf r)$ with $\vec{A}$ [Fig.~\ref{SupplementalMaterial}(a,d)], $\mathrm{Re}[A_x^2]$ [Fig.~\ref{SupplementalMaterial}(b,e)], and $\partial_xA_x$ [Fig.~\ref{SupplementalMaterial}(c,f)]. 
The spatial profile of $\mathrm{Re}[A_x^2]$ coincides with the induced amplitude modulation $\delta\psi(\mathbf r,t)$, 
while $\partial_xA_x$ matches the phase modulation $\delta\theta(\mathbf r,t)$ (See Fig.~2 in the main text).


